\documentclass[journal]{IEEEtran}

\usepackage{graphicx}
\usepackage{ulem}
\usepackage{amsmath}
\usepackage{array}
\newcommand{\PreserveBackslash}[1]{\let\temp=\\#1\let\\=\temp}
\newcolumntype{C}[1]{>{\PreserveBackslash\centering}p{#1}}
\newcolumntype{R}[1]{>{\PreserveBackslash\raggedleft}p{#1}}
\newcolumntype{L}[1]{>{\PreserveBackslash\raggedright}p{#1}}
\usepackage{colortbl,booktabs}
\usepackage{tabularx}
\usepackage{multicol}
\usepackage{booktabs}
\usepackage{array,color}
\usepackage[Symbol]{upgreek}
\usepackage{subfigure}
\usepackage{bm}
\usepackage{cite}
\usepackage{stfloats}
\usepackage{balance}
\usepackage{mathrsfs}
\usepackage{amsthm}
\usepackage{bbm}
\usepackage{amsmath, amssymb}
\newcommand{\tabincell}[2]{\begin{tabular}{@{}#1@{}}#2\end{tabular}}

\begin{document}

\bibliographystyle{IEEEtran}

\title{Compressive Sensing Techniques for Next-Generation Wireless Communications}

\author{Zhen Gao, Linglong Dai,~{\it Senior Member, IEEE}, Shuangfeng Han, Chih-Lin I,~{\it Senior Member, IEEE},
Zhaocheng Wang,~{\it Senior Member, IEEE}, and Lajos Hanzo,~{\it Fellow,~IEEE}%
\thanks{Z. Gao is with Advanced Research Institute for Multidisciplinary Science, Beijing Institute of Technology, Beijing 100081, P. R. China (E-mail: gaozhen16@bit.edu.cn).}
\thanks{L. Dai, and Z. Wang are with Tsinghua National Laboratory for
 Information Science and Technology (TNList), Department of Electronic Engineering,
 Tsinghua University, Beijing 100084, P. R. China (E-mails: \{daill, zcwang\}@tsinghua.edu.cn).} %
%
\vspace*{-6mm}
\thanks{S. Han and C. L. I are with Green Communication Research Center China Mobile Research Institute, Beijing 100053, P. R. China (E-mails:
\{hanshuangfeng,~icl\}@chinamobile.com).}
\thanks {L. Hanzo is with Electronics and Computer Science, University of Southampton, Southampton SO17 1BJ, U.K. (E-mail: lh@ecs.soton.ac.uk).}
}

\maketitle
\begin{abstract}
A range of efficient wireless processes and enabling techniques are
put under a magnifier glass in the quest for exploring different
manifestations of correlated processes, where sub-Nyquist sampling may
be invoked as an explicit benefit of having a sparse transform-domain
representation.  For example, wide-band next-generation systems
require a high Nyquist-sampling rate, but the channel impulse response
(CIR) will be very sparse at the high Nyquist frequency, given the low
number of reflected propagation paths. This motivates the employment
of compressive sensing based processing techniques for frugally
exploiting both the limited radio resources and the network
infrastructure as efficiently as possible. A diverse range of
sophisticated compressed sampling techniques is surveyed and we
conclude with a variety of promising research ideas related to
large-scale antenna arrays, non-orthogonal multiple access (NOMA), and
ultra-dense network (UDN) solutions, just to name a few.
\end{abstract}
\begin{IEEEkeywords}
5G, compressive sensing (CS), sparsity, massive MIMO, millimeter-wave (mmWave) communications, non-orthogonal multiple access (NOMA), ultra-dense networks (UDN). 
\end{IEEEkeywords}
\section{Introduction}\label{S1}



The explosive growth of traffic demand resulted in gradually approaching the system capacity of the
operational cellular networks \cite{5G}.
It is widely recognized that substantial system capacity improvement is required for 5G in the next decade \cite{5G}. To tackle this challenge, a suite of 5G techniques and
proposals have emerged, accompanied by: i) increased spectral efficiency relying on multi-antenna
techniques and novel multiple access techniques offering more bits/sec/Hz per node; ii) a larger transmission bandwidth relying on spectrum sharing and extension; iii) improved spectrum reuse relying
on network densification having more nodes per unit area.

Historically speaking, the transmission bandwidth has increased from 200 kHz in the 2G GSM system
to 5 MHz in the 3G, to at most 20 MHz in the 4G. Meanwhile, the number of antennas employed also
increases from 1 in the 2G/3G systems to 8 in 4G, along with the increasing density of both the base
stations (BSs) deployed and users supported. Despite the gradual quantitative increase of bandwidth,
number of antennas, density of BS and users, all previous wireless cellular networks have relied upon the
classic Nyquist sampling theorem, stating that any bandwidth-limited signal can be perfectly reconstructed,
when the sampling rate is higher than twice the signal's highest frequency.
However, the emerging
5G solutions will require at least 100 MHz bandwidth, hundreds of antennas, and ultra-densely deployed
BSs to support massive users. These qualitative changes indicate that applying Nyquist's sampling theorem to 5G techniques reminiscent of the previous 2G/3G/4G solutions may result in
unprecedented challenges: prohibitively large overheads, unaffordable complexity, and high cost and/or power consumption due to the large number of samples required. On the other hand,
compressive sensing (CS) offers a sub-Nyquist sampling approach to the reconstruction of sparse signals of an under-determined linear system in a computationally efficient manner~\cite{Eldar}. Given the
large bandwidth of next-generation systems and the proportionally high Nyquist-frequency, we arrive at an excessive number of resolvable multipath components, even though only a small fraction
of them is non-negligible. This phenomenon inspired us to sample the resultant sparse
channel impulse response (CIR) as well as other signals under the framework of CS, thus offering
us opportunities to tackle the above-mentioned challenges~\cite{Eldar}.
\begin{figure*}[tp]
\centering
\includegraphics[width=18cm]{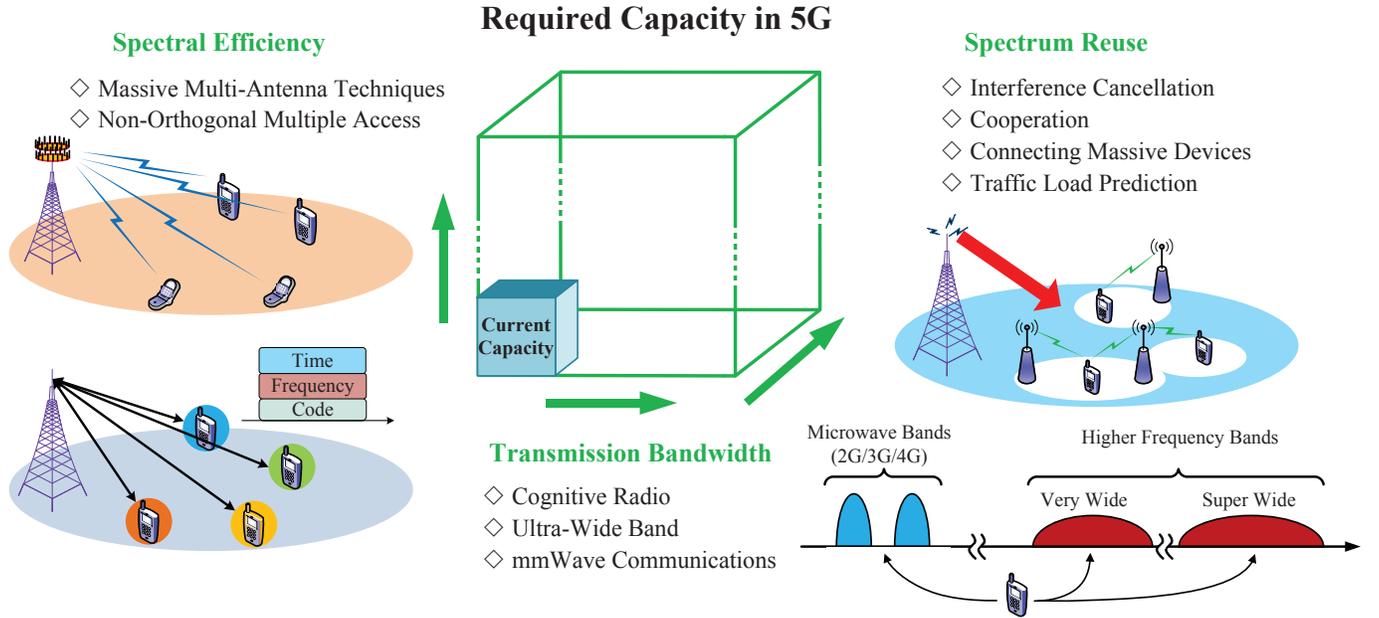}
\caption{Three promising technical directions for 5G.} \label{Cube}
\end{figure*}

To be more specific, in Section \ref{S2}, we first introduce the key
5G techniques, while in Section \ref{S1.1}, we present the concept of
CS, where three fundamental elements, four models, and the associated recovery algorithms are introduced.
Furthermore, in Sections \ref{Spectral_Efficiency},
\ref{Spectrum_Utilization} and \ref{HetNet}, we investigate the opportunities and challenges of applying the CS techniques to those key 5G solutions by exploiting the multifold sparsity inherent, as
briefly presented below:
\begin{itemize}
  \item We exploit the CIR-sparsity in the context of massive MIMO
    systems both for reducing the channel-sounding overhead required for
    reliable channel estimation, as well as the spatial modulation (SM)-based
    signal sparsity inherent in massive SM-MIMO and the codeword
    sparsity of non-orthogonal multiple access (NOMA) in order to
    reduce the signal detection complexity.
 \item We exploit the sparse spectrum occupation with the aid of
   cognitive radio (CR) techniques and the sparsity of the
   ultra-wide band (UWB) signal for reducing both the hardware cost as
   well as the power consumption.  Similarly, we exploit the CIR sparsity
   in millimeter-wave (mmWave) communications for improving the
   transmit precoding performance as well as for reducing the CIR
   estimation overhead.
  \item Finally, we exploit the sparsity of the interfering base
    stations (BSs) and of the traffic load in ultra-dense networks (UDN),
    where sparsity can be capitalized on by reducing the overheads
    required for inter-cell-interference (ICI) mitigation, for coordinated multiple points (CoMP) transmission/reception, for
    large-scale random access and for traffic prediction.
\end{itemize}

We believe that these typical examples can further inspire the
conception of a plethora of potential sparsity exploration and
exploitation techniques. Our hope is that you valued colleague might
also become inspired to contribute to this community-effort.

\section{Key Technical Directions in 5G}\label{S2}
The celebrated Shannon capacity formula indicates the total network capacity can be approximated as
\begin{equation}
{C_{{\rm{network}}}} \approx \sum\limits_i^I {\sum\limits_j^J {{W_{i,j}}{{\log }_2}\left( {1 + {\rho _{i,j}}} \right)} } ,
\end{equation}
where $i$ and $j$ are the indices of cells and channels, respectively, $I$ and $J$ are the numbers of cells and channels, respectively, $W_{i,j}$ and $\rho_{i,j}$ are the associated bandwidth and signal-to-interference-plus-noise ratio (SINR), respectively.
As shown in Fig. \ref{Cube} at a glance, increasing ${C_{{\rm{network}}}}$ for next-generation systems relies on 1) achieving an increased spectral efficiency with larger number of channels,
for example by spatial-multiplexing MIMO; 2) an increased transmission bandwidth including spectrum sharing and extension; and 3) better spectrum reuse relying on more cells per area 
for improving the area-spectral-efficiency (ASE).
To elaborate a little further:

1) Increased spectral efficiency can be achieved for example: first,~{massive multi-antenna aided spatial-multiplexing techniques} can
substantially boost the system capacity, albeit both the channel
estimation in massive MIMO \cite{massiveMIMO,my} and the signal
detection of massive spatial modulation (SM)-MIMO \cite{SM} remain
challenging issues; second,~{NOMA techniques are theoretically capable of
  supporting more users than conventional orthogonal multiple access
  (OMA) under the constraint of limited radio resources, but the
  optimal design of sparse codewords capable of approaching the NOMA
  capacity remains an open problem at the time of writing~\cite{SCMA}.

2) Larger transmission bandwidth may be invoked relying on both
{CR} \cite{CR,CR2} and UWB \cite{UWB,UWB0} techniques, both of which can
coexist with licenced services under the umbrella of spectrum sharing,
where the employment of sub-Nyquist sampling is of salient importance.
As another promising candidate, {mmWave communications} is capable of
facilitating high data rates with the aid of its wider bandwidth
\cite{5G,Hea_JSTSP,hybrid}.  However, due to the limited availability
of hardware at a low cost and owing to its high path-loss, both
channel estimation and transmit precoding are more challenging in
mmWave systems than those in the existing cellular systems.

{3) Better spectrum reuse can be realized with the aid of small cells \cite{5G}, which improves the ASE expressed in
  bits/sec/Hz/km$^2$.  However, how to realize interference
  mitigation, CoMP transmission/reception and massive random access imposes
  substantial challenges \cite{Interference,MRA,predict}.

\begin{table*}
\renewcommand{\arraystretch}{1.3}
\caption{Typical CS Models}
\label{tab:example}
\centering
\begin{tabular}{|c|c|c|c|c|}
  \hline
  \tabincell{c}{Types\\of Model}  &CS Models & Mathematical Expression & Illustration  \\
  \hline
  Model (1) & \tabincell{c}{Standard\\CS model \cite{Eldar}} & ${\bf{y}} = {\bf{\Phi}} {\bf{x}} ={\bf{\Phi \Psi s}} = {\bf{\Theta s}}$ & ${\bf{\Theta }} = {\bf{\Phi \Psi}}$ \\
  \hline
  Model (2) & \tabincell{c}{Signal separation by\\sparse representations \cite{Eldar}} & \tabincell{c}{${ \bf{\tilde y}} = \sum\limits_{p = 1}^P {{{\bf{\Theta }}_p}{{\bf{s}}_p}}  = {{\bf{\Theta }}_1}{{\bf{s}}_1} + \underbrace {\sum\limits_{p = 2}^P {{{\bf{\Theta }}_p}{{\bf{s}}_p}} }_{{\mathop{\rm int}} {\rm{erference}}}= {\bf{\Theta s}} $ \\${\bf{\Theta }}{\rm{ = }}\left[ {{{\bf{\Theta }}_1},{{\bf{\Theta }}_2}, \cdots ,{{\bf{\Theta }}_P}} \right]$, ${\bf{s}} = {\left[ {{\bf{s}}_1^{\rm{T}},{\bf{s}}_2^{\rm{T}}, \cdots ,{\bf{s}}_P^{\rm{T}}} \right]^{\rm{T}}}$}& \tabincell{c}{${\bf{s}}_p$\! and\! ${{\bf{\Theta }}_p}$\! are the $p$th sparse\\signal and the $p$th measure-\\ment matrix, respectively, $\bf{s}$\\is the sparse aggregate signal}  \\
  \hline
  Model (3) & \tabincell{c}{Block sparse signal \cite{Eldar}} & \tabincell{c}{${\bf{y}} = {\bf{\Theta s}}$, ${\bf{s}}$ appears the block sparsity, e.g., \\${\bf{s}}={[\underbrace {{s_1} \cdots {s_d}}_{{{\bf{s}}^{\rm{T}}}\left[ 1 \right]}\underbrace {{s_{d + 1}} \cdots {s_{2d}}}_{{{\bf{s}}^{\rm{T}}}\left[ 2 \right]} \cdots \underbrace {{s_{N - d + 1}} \cdots {s_N}}_{{{\bf{s}}^{\rm{T}}}\left[ L \right]}]^{\rm{T}}}$ }& \tabincell{c}{$dL\!=\!N$, and ${{{\bf{s}}^{\rm{T}}}\!\left[ l  \right]}$ for $1\le l \le L$\\has non-zero Euclidean norm for\\at most $k$ indices~~~~~~~~~~~~~~~~~~~} \\
  \hline
   Model (4) & \tabincell{c}{Multiple vector\\measurement (MMV) \cite{Eldar}}& \tabincell{c}{$\left[ {{{\bf{y}}_1},{{\bf{y}}_2}, \cdots ,{{\bf{y}}_P}} \right] = {\bf{\Theta }}\left[ {{{\bf{s}}_1},{{\bf{s}}_2}, \cdots ,{{\bf{s}}_P}} \right]$,\\$\left\{ {{{\bf{s}}_p}} \right\}_{p = 1}^P$ share the identical or partially \\common sparsity pattern~~~~~~~~~~~~~~~~~}& \tabincell{c}{$ {{{\bf{s}}_p}} $ and $ {{{\bf{y}}_p}} $ for $1\le p \le P$ are\\the sparse signal and measured\\signal~ associated with the ~$p$th\\observation, respectively~~~~~~~~~\\}\\
    \hline
\end{tabular}
\end{table*}

\section{Compressive Sensing Theory}\label{S1.1}
Naturally, most continuous signals from the real world exhibit some inherent redundancy or correlation,
which implies that the effective amount of information conveyed by them is typically lower than the
maximum amount carried by uncorrelated signals in the same bandwidth~\cite{Eldar}. This is exemplified by the inter-sample correlation of
so-called voiced speech segments, by adjacent video pixels, correlated fading channel envelopes, etc. Hence the number
of effective degrees of freedom of the corresponding sampled discrete time signals can be much smaller
than that potentially allowed by their dimensions. This indicates that these correlated time-domain (TD) signals typically can be represented by
much less samples in the frequency-domain (FD)~\cite{Eldar}, because correlated signals only have a few non-negligible low-frequency FD components.
Just to give a simple example,
a sinusoidal signal can be represented by a single non-zero
frequency-domain tone after the transformation by the Fast Fourier
transformation (FFT).  Sometimes this is also referred to as the
energy-compaction property of the FFT.  Against this background, CS
theory has been developed and applied in diverse fields, which shows
that the sparsity of a signal can indeed be exploited to recover a
replica of the original signal from fewer samples than that required
by the classic Nyquist sampling theorem.

To briefly introduce CS
theory, we consider the sparse signal ${\bf{x}} \in \mathbb{C}^{n
  \times 1}$ having the sparsity level of $k$ (i.e., $\bf{x}$ has only
$k\ll n$ non-zero elements), which is characterized by the measurement
matrix of ${\bf{\Phi}}\in \mathbb{C}^{m \times n}$ associated with $m
\ll n$, where ${\bf{y}} ={\bf{\Phi}} {\bf{x}}\in \mathbb{C}^{m \times
  1} $ is the measured signal.  In CS theory, the key issue is how to
recover $\bf{x}$ by solving the under-determined set of equations
${\bf{y}} ={\bf{\Phi}} {\bf{x}}$, given $\bf{y}$ and
$\bf{\Phi}$. Generally, $\bf{x}$ may not exhibit sparsity itself,
but it may exhibit sparsity in some transformed domain, which is
formulated as ${\bf{x}}= {\bf{\Psi}} {\bf{s}}$, where $ {\bf{\Psi}}$
is the transform matrix and $\bf{s}$ is the sparse signal associated
with the sparsity level $k$.  Hence we can formulate the standard CS
Model (1) of Table \ref{tab:example}.
Additionally, we can infer from the standard CS Model (1)
of Table I the equally important Models (2), (3), and (4) of
Table \ref{tab:example},
which can provide more reliable compression and recovery of sparse signals, when some of the specific sparse properties of practical applications are considered.
Specifically, Model (2) is capable of separating multiple sparse signals $\left\{ {{{\bf{s}}_p}} \right\}_{p = 1}^P$ associated with different
measurement matrices $\left\{ {{{\bf \Theta} _p}} \right\}_{p = 1}^P$ by recovering the aggregate sparse signal ${\bf{s}} = {\left[ {{\bf{s}}_1^{\rm{T}},{\bf{s}}_2^{\rm{T}}, \cdots ,{\bf{s}}_P^{\rm{T}}} \right]^{\rm{T}}}$;
Model (3) has the potential of improving the estimation performance of $\bf s$ by exploiting the block sparsity of $\bf s$, as shown in Table I; Model (4) is capable of enhancing the estimation performance of $P$ sparse
signals $\left\{ {{{\bf{s}}_p}} \right\}_{p = 1}^P$, when their identical/partially common sparsity pattern is exploited.

 Considering the standard
CS model, we arrive at the three fundamental elements of CS theory as follows.
1) {\it Sparse transformation} is essential for CS, since finding a suitable transform matrix $\bf{\Psi}$ can
  efficiently transform the original (non-sparse) signal $\bf{x}$ into the sparse signal $\bf{s}$. 
2) {\it Sparse signal compression} refers to the design of $\bf{\Phi}$ or ${\bf{\Theta }} = {\bf{\Phi \Psi }}$. $\bf{\Phi}$ should reduce the dimension of measurements,
      while minimizing the information loss imposed,
      which can be quantified in terms of
      the coherence or restricted isometry property (RIP) of $\bf{\Phi}$ or~${\bf{\Theta}}$ \cite{Eldar}.
3) {\it Sparse signal recovery algorithms} are important for the reliable reconstruction of $\bf{x}$ or $\bf{s}$ from the
  measured signal $\bf{y}$. 
Particularly, the CS algorithms widely applied in wireless communications can be mainly divided into three categories as follows.

i) {\it Convex relaxation algorithms} such as basis pursuit (BP) as well as BP de-noising (BPDN), and so on,
can formulate the CS problem as a convex optimization problem and solve them using convex optimization software like CVX~[2]. For instance, the CS problem for Model (1) of Table I
can be formulated as a Lagrangian relaxation of a quadratic program as
\begin{equation}\setcounter{equation}{2}
{\bf{\hat s}} = \arg \mathop {\min }\limits_{\bf{s}} {\left\| {\bf{s}} \right\|_1} + \lambda {\left\| {{\bf{y}} - {\bf{\Theta s}}} \right\|_2},
\end{equation}
with ${\left\| \cdot \right\|_1}$ and ${\left\|  \cdot \right\|_2}$ being ${l_1}$-norm and ${l_2}$-norm operators, respectively, and $\lambda>0$, and the resultant algorithms belong to the BPDN family. These algorithms usually require a small number of measurements, but they are complex, e.g., the complexity of BP algorithm is on the order of $O\left( {{m^2}{n^{3/2}}} \right)$~\cite{Eldar}.

ii) {\it Greedy iterative algorithms} can identify the support set in a greedy iterative manner. They have a low complexity and fast speed of recovery, but suffer from a performance loss, when the signals are not very sparse.
The representatives of these algorithms are orthogonal matching pursuit (OMP), CoSaMP, and subspace pursuit (SP), which have the complexity of $O\left( kmn \right)$~\cite{Eldar}.

iii) {\it Bayesian inference algorithms} like sparse Bayesian learning and approximate message passing infer the sparse unknown signal from the Bayesian viewpoint by considering the sparse priori. The complexity of these algorithms varies from individual to individual. For example, the complexity of Bayesian compressive sensing via belief propagation is $O\left( {n{{\log }^2}n} \right)$~\cite{Eldar}.
Note that, the algorithms mentioned above have to be further developed for Models (2)-(4) of Table I. For example, the group-sparse BPDN, the simultaneous OMP (SOMP), and the group-sparse Bayesian CS algorithms
tailored for MMV Model (4) are promising future candidates~\cite{Eldar}.

Since the conception of CS theory in 2004, it has been extensively developed, extended and applied to practical systems. Indeed, prototypes for MIMO radar, CR, UWB, and so on based on CS
theory have been reported by Eldar's research group~\cite{Eldar}.
Undoubtedly, the emerging CS theory 
provides us with a revolutionary tool for reconstructing
signals, despite using sub-Nyquist sampling rates \cite{Eldar}.
Therefore, how to exploit CS theory in the emerging 5G wireless
networks has become a hot research topic
\cite{massiveMIMO,my,SM,CR,CR2,UWB,hybrid,Hea_JSTSP,Interference,MRA,predict,UWB0}.
By exploring and exploiting the inherent sparsity in all aspects of
wireless networks, we can create more efficient 5G networks.  In the
following sections, we will explore and exploit the sparsity inherent
in future 5G wireless networks in the context of the three specific
technical directions discussed in Section~\ref{S2}.

\section{Higher Spectral Efficiency}\label{Spectral_Efficiency}

The first technical direction to support the future 5G vision is to
increase the spectral efficiency, where massive MIMO, massive
SM-MIMO and NOMA schemes constitute promising candidates.  This
section will discuss how to explore and exploit the sparsity inherent
in these key 5G techniques.
\subsection{Massive MIMO Schemes}\label{Massive_MIMO}
Massive MIMO employing hundreds of antennas at the BS are capable of
simultaneously serving multiple users at an improved spectral- and the
energy-efficiency \cite{massiveMIMO,my}.  Although massive MIMO
indeed exhibit attractive advantages, a challenging issue that hinders
the evolution from the current frequency division duplex (FDD)
cellular networks to FDD massive MIMO is the indispensible estimation
and feedback of the downlink FDD channels to the transmitter.
However, for FDD massive MIMO, the users have to estimate the
downlink channels associated with hundreds of transmit and receive
antenna pairs, which results in a prohibitively high pilot overhead.
Moreover, even if the users have succeeded in acquiring accurate
downlink channel state information (CSI), its feedback to the BS
requires a high feedback rate. Hence the codebook-based
CSI-quantization and feedback remains challenging, while the overhead
of analog CSI feedback is simply unaffordable~\cite{my}.  By contrast,
in time division duplex (TDD) massive MIMO, the downlink CSI can be
acquired from the uplink CSI by exploiting the channel's reciprocity,
provided that the interference is also similar at both ends of the
link. Furthermore, the pilot contamination may significantly degrade
the system's performance due to the limited number of orthogonal
pilots, which hence have to be reused in adjacent cells~\cite{massiveMIMO}.

Fortunately, recent experiments have shown that due to the limited
number of significant scatterers in the propagation environments and
owing to the strong spatial correlation inherent in the co-located
antennas at the BS, the massive MIMO channels exhibit sparsity either
in the delay domain~\cite{massiveMIMO} or in the angular domain or in
both~\cite{my}. For massive MIMO channels observed in the delay
domain, the number of paths containing the majority of the received
energy is usually much smaller than the total number of CIR taps,
which implies that the massive MIMO CIRs exhibit sparsity in the delay
domain and can be estimated using the standard CS Model (1) of Table I, where
$\bf s$ is the sparse delay-domain CIR, $\bf \Theta$ consists of pilot signals,
and $\bf y$ is the received signal~\cite{massiveMIMO}.
Due to the co-located nature of the antenna elements, the
CIRs associated with different transmit and receiver antenna pairs further exhibit structured sparsity, which manifests
itself in the block-sparsity Model (3) of \cite{massiveMIMO}.
Moreover, the BS antennas are usually
found at elevated location with much few scatterers around, while the
users roam at ground-level and experience rich scatterers.  Therefore,
the massive MIMO CIRs seen from the BS exhibit only limited angular
spread, which indicates that the CIRs exhibit sparsity in the angular
domain~\cite{my}.
Due to the common scatterers shared by multiple users close to each other, the
massive multi-user MIMO channels further have the structured sparsity and can be jointly estimated using the
MMV Model (4) of Table I~\cite{my}. Additionally, this sparsity can also be
exploited for mitigating the pilot contamination in TDD massive MIMO,
where the CSI of the adjacent cells can be estimated with the aid of
the signal separation Model (2) for further interference mitigation or
for multi-point cooperation.

{\it Remark:} Exploiting the sparsity of massive MIMO channels
with the aid of CS theory to reduce the overhead required for channel
estimation and feedback are expected to solve various open challenges
and constitute a hot topic in the field of massive
MIMO~\cite{massiveMIMO,my}.  However, if the pilot signals
of CS-based solutions are tailored to a sub-Nyquist sampling
rate, ensuring its compatibility with the existing systems based on the
classic Nyquist sampling rate requires further research.

\begin{figure}[tp]
\centering
\includegraphics[width=9cm]{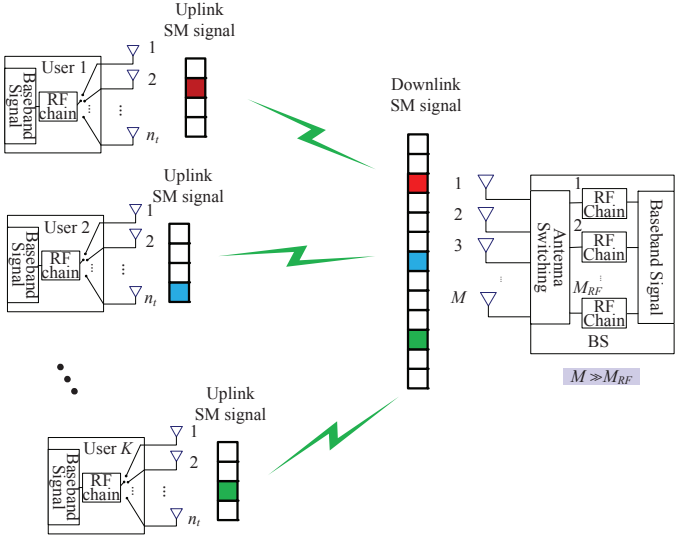}
\caption{The SM signals in massive SM-MIMO systems are sparse.} \label{fig:MAT}
\end{figure}
\subsection{Massive SM-MIMO Schemes}\label{SM_MIMO}

In massive MIMO systems, each antenna requires a dedicated radio
frequency (RF) chain, which will substantially increase the power
consumption of RF circuits, when the number of BS antennas becomes
large.  To circumvent this issue, as shown in Fig. \ref{fig:MAT}, the
BS of massive SM-MIMO employs hundreds of antennas, but a much
smaller number of RF chains and antennas is activated for
transmission.  Explicitly, only a small fraction of the antennas is
selected for the transmission of classic modulated signals in each
time slot.  For massive SM-MIMO, a 3-D constellation diagram
including the classic signal constellation and the spatial
constellation is exploited.  Moreover, massive SM-MIMO can also be
used in the uplink \cite{SM}, where multiple users equipped with a
single-RF chain, but multiple antennas can simultaneously transmit
their SM signals to the BS. In this way, the uplink throughput can
also be improved by using SM, albeit at the cost of having no transmit
diversity gain. This problem can be mitigated by activating a limited
fraction of the antennas.

Due to the potentially higher number of transmit antennas than the
number of activated receive antennas, signal detection and channel
estimation in massive SM-MIMO can be a large-scale under-determined
problem.  The family of optimal maximum likelihood or near-optimal
sphere decoding algorithms suffers from a potentially excessive
complexity.  By contrast, the conventional low-complexity linear
algorithms, such as the linear minimum mean square error (LMMSE)
algorithm, suffer from the obvious performance loss inflicted by
under-determined rank-deficient systems.  Fortunately, it can be
observed that in the downlink of massive SM-MIMO, since only a
fraction of the transmit antennas are active in each time slot, the
downlink SM signals are sparse in the signal domain. Hence, we can use
the standard CS Model (1) of Table I for developing SM signal
detection, where
$\bf{s}$ is the sparse SM signal, $\bf {\Theta}$ is the MIMO channel matrix, and
$\bf y$ is the received signal. Moreover, observe in Fig. 2 that for the uplink of massive
SM-MIMO, each user's uplink SM signal also exhibits sparsity, thus the
aggregated SM signal incorporating all of the multiple users' uplink
SM signals exhibits sparsity.  Therefore, it is expected
that by exploiting the sparsity of the aggregated SM
signals, we can use the signal separation Model (2) of Table I to
develop a low-complexity, high-accuracy signal detector for improved
uplink signal detection~\cite{SM}.

{\it Remark:}
The sparsity of SM signals can be exploited for reducing the computational complexity of signal detection at the receiver.
To elaborate a little further, channel estimation in massive SM-MIMO
is more challenging than that in massive MIMO, since only a fraction
of the antennas are active in each time slot.  Hence, how to further
explore the intrinsic sparsity of massive SM-MIMO channels and how to
exploit the estimated CSI associated with the active antennas to
reconstruct the complete CSI is a challenging problem requiring
further investigations.

\begin{figure}[tp]
\centering
\includegraphics[width=8.5cm]{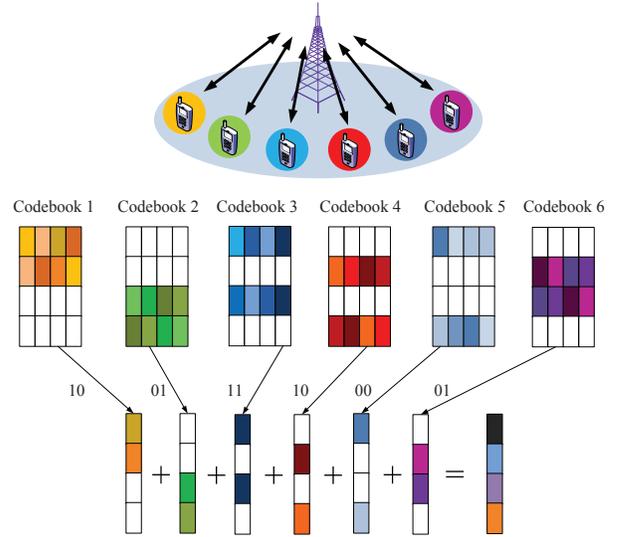}
\caption{SCMA is capable of supporting overloaded transmission by sparse code domain multiplexing.} \label{fig:SCMA}
\end{figure}

\subsection{Sparse Codewords in NOMA systems}\label{SCMA}
Cellular networks of the first four generations obeyed different
orthogonal multiple access (OMA) techniques~\cite{SCMA}.  In contrast
to conventional OMA techniques, such as frequency division multiple
access (FDMA), time division multiple access (TDMA), and orthogonal
frequency division multiple access (OFDMA), NOMA systems are
potentially capable of supporting more users/devices by using
non-orthogonal resources, albeit typically at the cost of increased
receiver complexity.

As a competitive NOMA candidate, sparse code multiple access (SCMA)
supports the users with the aid of their unique user-specific
spreading sequence, which are however non-orthogonal to each other -
similar to classic m-sequences, as illustrated in
Fig. \ref{fig:SCMA}~\cite{SCMA}.  Each codeword exhibits sparsity and
represents a spread transmission layer.  In the uplink, the BS can
then uniquely and unambiguously distinguish the different sparse
codewords of multiple users relying on non-orthogonal resources.  In
the downlink, more than one transmission layers can be transmitted to
each of the multiple users with the aid of the above-mentioned
non-orthogonal codewords.
The SCMA signal detection problem can be readily formulated as the signal
separation Model (2) of Table I, where the columns of ${{{\bf \Theta} _p}} $
consist of the $p$th user's codewords, and ${\bf{s}}_p$ is a vector with 0 and 1 binary values and only one non-zero element.
 Amongst others, the low-complexity
message passing algorithm (MPA) can be invoked by the receiver for
achieving a near-maximum-likelihood multi-user detection performance.

{\it Remark:} The optimal codeword design of SCMA and the associated
multi-user detector may be designed with the aid of
CS theory for improving the performance versus complexity trade-off~\cite{SCMA}. 

\section{Larger Transmission Bandwidth}\label{Spectrum_Utilization}
The second technical direction contributing to the 5G vision is based
on the larger transmission bandwidth, where the family of promising
techniques includes CR, UWB and mmWave communication.  How we might
explore and exploit sparsity in these key 5G techniques will be
addressed in this section.

\subsection{Cognitive Radio}\label{CR}
It has been revealed in the open literature that large portions of the
licensed spectrum remains under-utilized 
\cite{CR,CR2}, since the licensed users
may not be fully deployed across the licensed territory or might not
occupy the licensed spectrum all the time, and guard bands may be
adopted by primary users (PUs).  Due to the sparse spectrum exploitation, 
CR has been advocated for dynamically sensing the
unused spectrum and for allowing the secondary users (SUs) to exploit the
spectrum holes, while imposing only negligible interference on the PUs.

However, enabling dynamic spectrum sensing and sharing of the entire
spectral bandwidth is challenging, due to the 
high Nyquist
sampling rate for SUs to 
sense
a broad spectrum. 
To exploit the
low spectrum occupancy by the licensed activities, as verified by
extensive experiments and field tests~\cite{CR}, the compressive
spectrum sensing concept, which can be described by the standard CS
Model (1) of Table I, has been invoked for sensing the spectrum at
sub-Nyquist sampling rates.  In CR networks, every SU can sense the
spectrum holes, despite using a sub-Nyquist sampling rate.  However,
this strategy may be susceptible to channel fading, hence
collaborative sensing relying on either centralized or distributed
processing has also been proposed \cite{CR,CR2}. Due to the collaborative
strategy, the sparse spectrum seized by each SU may share common
components, which can be readily described by the MMV Model (4) of
Table I to achieve spatial diversity~\cite{CR}.
Moreover, integrating a geo-location
database into compressive CR is capable of further improving the performance attained~\cite{CR2}.

{\it Remark:} CS-based CR can facilitate the employment of
low-speed analog-to-digital-converter (ADC) instead of the high-speed
ADC required by conventional Nyquist sampling theory.  In closing we
mention that in addition to sensing the spectrum holes by conventional
CR schemes, Xampling is also capable of demodulating the compressed
received signals, provided that their transmission parameters, such as
their frame structure and modulation modes are known~\cite{Eldar}.

\subsection{Ultra-Wide Band Transmission}\label{UWB}
UWB systems are capable of achieving Gbps data rates in short range
transmission at a low power consumption~\cite{UWB,UWB0}. Due to the ultra-wide
bandwidth utilized at a low power-density, UWB may
coexist with licenced services relying on frequency
overlay. Meanwhile, the ultra-short duration of time-hopping UWB
pulses enables it to enjoy fine time-resolution and multipath
immunity, which can be used for wireless location.

According to Nyquist's sampling theorem, the GHz bandwidth of UWB
signals requires a very high Nyquist sampling rate, which leads to the
requirement of high-speed ADC and to the
associated strict timing control at the receiver. This increases both
the power consumption and the hardware cost.  However, the intrinsic
time-domain sparsity of the received line-of-sight (LOS) or
non-line-of-sight (NLOS) UWB signals inspires the employment of an
efficient sampling approach under the framework of CS, where
the sparse UWB signals can be recovered by using sub-Nyquist sampling
rates. Moreover, the UWB signals received over multipath channels can
also be approximately considered as a linear combination of several
signal bases, as in the standard CS Model (1) of Table I, where these
signal bases are closely related to the UWB waveform, such as the
Gaussian pulse or it derivatives \cite{UWB,UWB0}.
Compared to those users, who only
exploit the time-domain sparsity of UWB signals, the latter approach
can lead to a higher energy-concentration and to the further
improvement of the sparse representation of the received UWB signals,
hence enhancing the reconstruction performance of the UWB signals
received by using fewer measurements.
Besides, CS can be further applied to
estimate channels in UWB transmission by formulating it as MMV Model (4) of Table 1, where
the common sparsity of multiple received pilot signals is exploited \cite{UWB}.

{\it Remark:} The sparsity of the UWB signals facilitates the reconstruction of the UWB signals from observations sampled by the low-speed and power-saving ADCs relying on sub-Nyquist sampling. 
The key challenge is how to extract
the complete information characterizing the
analog UWB signals from the compressed measurements.
Naturally, if the receiver only wants to extract the information conveyed by the
UWB signals, it may be capable of
directly processing the compressed measurements by
skipping the reconstruction of the UWB signals \cite{UWB0}.


\subsection{Millimeter-Wave Communications}\label{mmWave}
The crowded microwave frequency band and
the growing demand for increased data rates
motivated researchers to reconsider the under-utilized
mmWave spectrum (30$\sim$100 GHz). 
Compared to existing cellular communications operating
at sub-6 GHz frequencies,
mmWave communications have three
distinctive features:
a) the spatial sparsity of channels due to the high path-loss
 of NLOS paths, b) the low signal-to-noise-ratio (SNR)
 experienced before beamforming, and c)
 the much smaller number of RF chains than that of the
 antennas due to the hardware constraints in mmWave communications \cite{Hea_JSTSP,hybrid}.
Hence, the spatial sparsity of channels can be readily exploited
for designing cost-efficient mmWave communications.

\subsubsection{Hybrid Analog-Digital Precoding}\label{Sparse_Precoding}
The employment of transmit precoding is important for mmWave MIMO
systems to achieve a large beamforming gain for the sake of
compensating their high pathloss.  However, the practical hardware
constraint makes the conventional full-digital precoding in mmWave
communications unrealistic, since a specific RF chain required by each
antenna in full-digital precoding may lead to an unaffordable hardware
cost and to excessive power consumption. Meanwhile, conventional
analog beamforming is limited to single-stream transmission and hence
fails to effectively harness spatial multiplexing. To this end,
hybrid analog-digital precoding relying on a much lower number of RF
chains than that of the antennas has been proposed, where the
phase-shifter network can be used for partial beamforming in the
analog RF domain for the sake of an improved spatial
multiplexing~\cite{hybrid}.

The optimal array weight vectors of analog precoding can be selected
from a set of beamforming vectors prestored according to the estimated
channels. Due to the limited number of RF chains and as a benefit of
spatial sparsity of the mmWave MIMO channels, the hybrid precoding can
be formulated as a sparse signal recovery problem, which was
referred to as spatially sparse precoding [Equ. (18) in 11]. This problem
can be efficiently solved by the modified OMP algorithm. However, the operation of this CS-based hybrid precoding
scheme is limited to narrow-band channels, while practical broadband
mmWave channels exhibit frequency-selective fading, which leads to a
frequency-dependent hybrid precoding across the bandwidth
\cite{hybrid}. For practical dispersive channels where OFDM is
likely to be used, it is attractive to design different digital
precoding/combining matrices for the different subchannels,
which may then be combined with a common
analog precodering/combining matrix with the aid of CS theory.
%

\subsubsection{Channel Estimation}\label{CE}
Hybrid precoding relies on accurate channel estimation, which is
practically challenging for mmWave communications relying on
sophisticated transceiver algorithms, such as multiuser MIMO
techniques.  In order to reduce the training overhead required for
accurate channel estimation, CS-based estimation schemes have been
proposed in~\cite{Hea_JSTSP,hybrid} by exploiting the sparsity of
mmWave channels.  Compared to conventional MIMO systems, channel
estimation designed for mmWave massive MIMO in conjunction with
hybrid precoding can be more challenging due to the much smaller
number of RF chains than that of the antennas.  The mmWave massive
MIMO flat-fading channel estimation can be formulated as the standard CS Model (1)
of Table I [Equ. (24) in 11], where $\bf s$ is the sparse channel vector in the angular domain, the hybrid precoding and combining
matrices  as well as the angular domain transform matrix compose $\bf \Theta$. 
While for dispersive mmWave MIMO channels,
the sparsity of angle of arrival (AoA), angle of departure (AoD), and multipath delay indicates that the
channel has a low-rank property. This property can be leveraged to reconstruct
the dispersive mmWave MIMO CIR, despite using a reduced number of observations \cite{Hea_JSTSP}.

{\it Remark:} By exploiting the sparsity of mmWave channels, CS can be
readily exploited both for reducing the complexity of hybrid precoding
and for mitigating the training overhead of channel estimation.
However, as to how we can extend the existing CS-based solutions from
narrow-band systems to broadband mmWave MIMO systems is still under
investigation.

\section{Better Spectrum Reuse}\label{HetNet}
The third technical direction to realize the 5G vision is to improve
the frequency reuse, which can be most dramatically improved by
reducing the cell-size \cite{5G}.  Ultra-dense small cells
including femocells, picocells, visible-light atto-cells are capable
of supporting seamless coverage, in a high energy efficiency, and a
high user-capacity. Explicitly, they can substantially decrease the
power consumption used for radio access, since the shorter distance
between the small-cell BSs and the users reduces the path-loss
\cite{Interference,MRA,predict}.  This section will address how to
explore and exploit the sparsity in dense networks under the framework
of CS theory.

\begin{figure*}[!t]
     \centering
     \includegraphics[width=18cm, keepaspectratio]
     {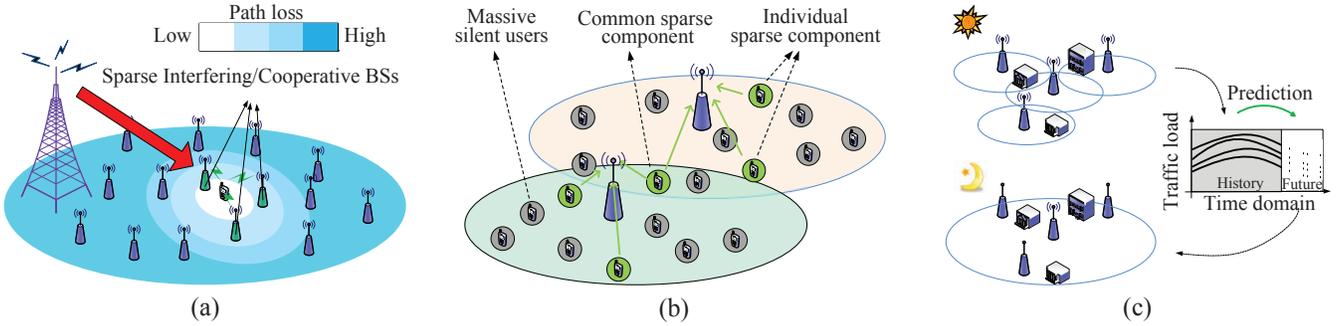}
    \caption{Sparsity in ultra-dense networks: (a) Sparse interfering BSs; (b) Sparsity of active users can be exploited to reduce the overhead for massive random access; (c) Low-rank property of large-scale traffic matrix facilitates its reconstruction
      with reduced overhead to dynamically manage the network.}
     \label{fig:HetNet}
\end{figure*}
\subsection{BSs Identification}\label{Coordination}
The ultra-dense small cells may impose non-negligible ICI, which
significantly degrades the received
SINR. Thus, efficient
interference cancellation is required for such interference-limited
systems.  In conventional cellular systems, orthogonal time-,
frequency-, and code resources can be used for effectively mitigating
the ICI.  By contrast, in the ultra-dense small cells, mitigating the
ICI in the face of limited orthogonal resources remains an open
challenge \cite{Interference}.

In the ultra-dense small cells of Fig. \ref{fig:HetNet} (a), a user will be interfered by multiple interfering BSs. The actual number of interfering BSs for a certain user is usually small, although the number of available BSs can be large. Hence, the identification of the interfering BSs can be formulated based on
 the CS Model (1) of Table I, where indices of non-zero elements in $\bf s$ corresponds to the interfering BSs, $\bf \Theta$ consists of the training signals, and $\bf y$ is the received signal at the users.
To identify the interfering BSs, each BS transmits non-orthogonal training signals based on their respective cell identity.
Then the users detect both the identity and even the CSI of the interfering BSs from the non-orthogonal received signals at a small overhead.
Moreover, the ICI may be further mitigated by using the signal separation Model (2) of Table I via the CoMP transmission, where the interfering BSs becomes the coordinated BSs. 
Additionally, by exploiting the observations from multiple antennas in the spatial domain and/or frames in the temporal domain, the block-sparsity Model (3) and MMV Model (4) of Table I can be further considered for improved performance~\cite{Interference}.


{\it Remark:} The identification of the interfering BSs can be formulated
as a CS problem for reducing the associated overhead, by exploiting the fact that the actual number of BSs interfering with
a certain user's reception is usually smaller than the total number of BSs.
Under the framework of CS, designing optimal non-orthogonal training signals and robust yet low-complexity detection algorithms for identifying these potential BSs with limited resources are still under investigation.
\subsection{Massive Random Access}\label{MAC}
It is a widely maintained consensus that the Internet of Things (IoT)
will lead to a plethora of devices connecting to dense networks for
cloud services in 5G networks.  However, the conventional orthogonal
resources used for multiple access impose a hard user-load limit,
which may not be able to cope with the massive connectivity ranging
from $10^2/\rm{km}^2$ to $10^7/\rm{km}^2$ for the IoT~\cite{5G}.  It
can be observed for each small-cell BS that although the number of
potential users in the coverage area can be large, the proportion of
active users in each time-slot is likely to remain small due to the
random call initiation attempts of the users accessing typical bursty
data services, as shown in Fig. \ref{fig:HetNet} (b)~\cite{MRA}.  In
this article, this phenomenon is referred to as the sparsity of
traffic, which points in the direction of CS-based massive random
access.

More particularly, in the uplink, the users transmit their unique
non-orthogonal training signals to access the cellular networks.  As a
result, the small-cell BSs have to detect multiple active users based
on the limited non-orthogonal resources. This multi-user detection
process can be described by the signal separation Model (2) of Table
I [Equ. (2) in 14].  Moreover, due to the ultra-dense nature of the small cells, the
adjacent small-cell BSs can also receive some common signals, which
implies that the adjacent small-cell BSs may share some common sparse
components. Were considering small-cell BSs constituted by the remote radio head (RRH) of the cloud radio access network (C-RAN) architecture, 
this sparse active-user detection carried out at the baseband unit (BBU) can be characterized by the MMV
Model (4) of Table I. By exploiting this structured sparsity, it is
expected that further improved active user detection performance can
be achieved.

{\it Remark:} The sparsity of traffic in UDN can be exploited for mitigating the access overhead with the aid of CS theory.
Compared to the identification of BSs in the downlink,
supporting large-scale random access in the uplink is more
challenging: 1) Since the number of users is much higher than that of
the BSs, the design of non-orthogonal training signals under the CS
framework may become more difficult; 2) The centralized cooperative
processing may be optimal, but the compression of the feedback
required for centralized processing may not be trivial; 3) Distributed
processing contributes an alternative technique of reducing the
feedback overhead, but the design of efficient CS algorithms remains
challenging.

\subsection{Traffic Estimation and Prediction for Energy-Efficient Dense Networks}\label{Traffic_prediction}
It has been demonstrated that the majority of power consumption for
the radio access is dissipated by the BSs, but this issue is more
challenging in dense networks \cite{5G}.  To dynamically manage the
radio access for the sake of improved energy efficiency, the
estimation of traffic load is necessary. However, under the classic
Nyquist sampling framework,
to estimate the large-scale traffic matrix for UDN, 
the measurements required as well as the associated storage, feedback, and energy consumption may become prohibitively
high.  Therefore, it is necessary to explore efficient techniques of estimating the traffic load for dense networks.

Experiments have shown that the demand for radio access exhibits the
obvious periodic variation on a daily basis and it also has a spatial
variation due to human activities \cite{{5G}}. 
The strong spatio-temporal
correlation of traffic load indicates that the indicator matrix of
traffic load exhibits a low rank, which inspires us to reconstruct the
complete indicator matrix with the aid of sub-Nyquist sampling
techniques~\cite{Eldar}.
When partial traffic data is missing, a spatio-temporal Kronecker compressive sensing
method may be involved for recovering the traffic matrix as the standard CS Model (1) of Table I [Equ. (15) in 15].
This may motivate us to exploit the low-rank property for estimating the complete traffic matrix with a reduced number of observations.
Furthermore, if the past history of
the traffic load has been acquired, traffic prediction may be obtained
by exploiting the low-rank property of the indicator matrix, and then the BSs can dynamically
manage the network for the improved energy efficiency. 
This process is illustrated in
Fig. \ref{fig:HetNet}~(c).
To achieve global traffic prediction from the different BSs, the
estimate of traffic load sampled by different sensors has to be fed
back to the fusion center, which may impose a huge overhead.  This
challenge may be mitigated by using part of the historic data for
traffic prediction and by exploiting the low-rank nature of the
indicator matrix. Moreover, since the spatial correlation of traffic
load is reduced as a function of the distance of different BSs, using
distributed CS-based traffic prediction with limited feedback may
become a promising alterative approach to be further studied.

{\it Remark:} The low-rank nature of the traffic-indicator matrix can
be exploited for reconstructing the complete indicator matrix with the aid of sub-Nyquist sampling techniques. In this way,
the measurements used for traffic prediction or their feedback to the fusion center can be reduced.
\section{Conclusions}\label{Conclusions}
CS has inspired the entire signal processing community and in this
treatise we revisited the realms of next-generation wireless communications technologies.
On the one hand, the very wide bandwidth, hundreds of antennas, and ultra-densely deployed
BSs to support massive users in those 5G techniques will result in the prohibitively large overheads, unaffordable complexity,
high cost and/or power consumption due to the large number of samples required by Nyquist sampling theorem.
On the other hand, CS theory has provided a sub-Nyquist sampling approach to efficiently tackle the above-mentioned challenges for these key 5G techniques.
We have investigated the
exploitation of sparsity in key 5G techniques from three technical
directions and four typical models.  Furthermore, we have discussed a
range of open problems and future research directions from the
perspective of CS theory.
The theoretical research on CS-based next generation
communication technologies has made substantial progress, but its applications in practical systems
still have to be further investigated. CS algorithms exhibiting reduced complexity and increased reliability, as well as compatibility with the current systems and hardware platforms constitute promising
potential future directions.
It may be anticipated that
CS will play a critical role in the design of future wireless
networks.

Hence our hope is that you valued colleague might like to
join this community-effort.


\end{document}